\documentclass[twocolumn, prc,superscriptaddress,showpacs,amsmath,amssymb]{revtex4}

\usepackage{graphicx}
\usepackage{dcolumn}
\usepackage{bm}
\usepackage{CJK}
\usepackage{txfonts}
\usepackage{color}
\usepackage{tabularx}
\usepackage{hyperref}

\def\dfrac#1#2{{\displaystyle\frac{#1}{#2}}}

\def\beq{\begin{equation}}
\def\eeq{\end{equation}}
\def\bea{\begin{eqnarray}}
\def\eea{\end{eqnarray}}

\def\fun#1#2{\lower3.6pt\vbox{\baselineskip0pt\lineskip.9pt
  \ialign{$\mathsurround=0pt#1\hfil##\hfil$\crcr#2\crcr\sim\crcr}}}

\usepackage{ulem}
\newcommand{\delete}{\bgroup\markoverwith{\textcolor{red}{\rule[0.5ex]{2pt}{1pt}}}\ULon}

\begin{document} 

\title{Proton radioactivity described by covariant density functional theory with Similarity Renormalization Group method}

\author{Qiang Zhao}
\affiliation{School of Nuclear Science and Technology, Lanzhou University, Lanzhou 730000, China}
\author{Jian Min Dong}
\affiliation{Research Center for Nuclear Science and Technology, Lanzhou University and Institute of Modern Physics of CAS, Lanzhou 730000, China}
\author{Jun Ling Song}
\affiliation{School of Nuclear Science and Technology, Lanzhou University, Lanzhou 730000, China}
\author{Wen Hui Long}\email{longwh@lzu.edu.cn}
\affiliation{School of Nuclear Science and Technology, Lanzhou University, Lanzhou 730000, China}

\begin{abstract}
  Half-life of proton radioactivity of spherical proton emitters is studied within the scheme of covariant density functional (CDF) theory, and for the first time the potential barrier that prevents the emitted proton is extracted with the similarity renormalization group (SRG) method, in which the spin-orbit potential along with the others that turn out to be non-negligible can be derived automatically. The spectroscopic factor that is significant is also extracted from the CDF calculations. The estimated half-lives are found in good agreement with the experimental values, which not only confirms the validity of the CDF theory in describing the proton-rich nuclei, but also indicates the prediction power of present approach to calculate the half-lives and in turn to extract the structural information of proton emitters.
\end{abstract}


\maketitle
 
With continuous development of the radioactive ion beam facilities, the exotic nuclei far away from the $\beta$-stability line attract extensive interests for the new phenomena they present. One of the typical representatives is the proton radioactivity at the vicinity of proton drip line, firstly observed in an isomeric state of $^{53}$Co in 1970 \cite{Co1,Co2}. Since then more and more proton emitters ranging from $Z = 51$ to $83$ have been identified with nuclear ground states or isomeric states \cite{Sonzogni2002}. Essentially, it is significant to study the proton emission which corresponds to the fundamental existence limits of neutron-deficient nuclei, i.e., the proton drip line, and it also can be treated as the inverse reaction of the rapid proton capture process that plays an important role in understanding the origin of the elements in the universe \cite{Wallace1981}. Moreover specific aspects of nucleonic interactions could be isolated and amplified in the proton emitters due to their extreme proton excess \cite{Blank2008}. In particular combined with theoretical analysis, nuclear structural information can be extracted from measurements of half-life, proton branching ratio (fine structure), the energy and angular momentum transfer $l$ carried away by the emitted proton, etc. The fact that the half-life of proton emission is sensitive to the $Q$-value and angular momentum transfer $l$, not only helps to determine the orbit of the emitted proton in parent nucleus in experiments, but also provides an efficient way to test theoretical models in exploring the neutron-deficient nuclear systems.

Theoretically various methods have been employed in describing the properties of proton emitters, such as the spectroscopic factor and the half-life (for review see Ref. \cite{Delion2006}). For the half-life that can be measured experimentally, a semiclassical method is applied by treating the proton emission as quantum tunneling through a potential barrier, which is composed of the Coulomb repulsion, centrifugal barrier and effective nuclear potential. Several approaches have been employed in constructing the effective nuclear potential, e.g., in terms of the density-dependent M3Y effective interaction \cite{Basu2005}, the effective interaction of Jeukenne, Lejeume, and Mahaux \cite{Bhattacharya2007}, the renormalized M3Y effective interaction \cite{Qian2010}, the R3Y interaction \cite{Sahu2011}, the finite-range effective interaction of Yukawa form \cite{Routray2011}, the Skyrme interactions \cite{Routray2012}, and also those from phenomenological unified fission model \cite{Balasubramaniam2005} and generalized liquid drop model \cite{Dong2009,Zhang2010}. In present work, the potential barriers are constructed under the scheme of covariant density functional (CDF) theory \cite{Walecka1974,Serot1986} with an alterative method.

Over the past years, the CDF theory based on the meson exchange diagram of nuclear force, in which the self-consistent treatment of the spin-orbit interaction is guaranteed by the covariant structure of the theory itself \cite{Furnstahl2004}, has attracted much attention for its great success in describing the structures of stable nuclei, neutron-rich nuclei, proton-rich nuclei, super-deformed nuclei and super-heavy nuclei \cite{Gambhir1990, Ring1996, Vretenar2005, Meng2006, Niksic2011}. There also exist some investigations on the properties of proton emitters within the CDF scheme \cite{Lalazissis1999-1, Lalazissis1999-2, Vretenar1999, Geng2004, Yao2008} and good agreements with the experimental data are achieved on the single proton separation energy and other relevant quantities. Additionally appropriate descriptions on the half-lives of proton radioactivity were provided by Sahu et al. \cite{Sahu2011} and Ferreira et al. \cite{Ferreira2011} using the CDF model.

In this study, we present a full calculation of half-lives of proton radioactivity within the CDF scheme and the potential barriers of the proton emitters are constructed with the similarity renormalization group (SRG) method \cite{Wegner1994, Guo2012, Guo2014} for the first time. Specifically, to be compatible with the WKB approximation in calculating the half-life, the Dirac equation is reduced into non-relativistic Schr\"odinger-type equations by the SRG approach, and it leads to diagonalized single-particle Hamiltonian and decoupled upper and lower components of the spinors \cite{Guo2012, Liang2013, Shen2013}. Namely, the Dirac equation is transferred into two independent Schr\"odinger-type equations respectively for the upper and lower components, and the potential from the upper one that describes nucleons in the Fermi sea is what we need, in which the spin-orbit potential along with other corrections can be identified explicitly without additional free parameters. Particularly, the spectroscopic factor that reflects the important information of nuclear structure is also taken into account and is calculated under the CDF scheme combined with BCS pairing treatment (CDF + BCS) \cite{Dong2009}.


Starting from an effective CDF Lagrangian containing the degrees of freedom associated with nucleon ($\psi$), mesons (the isoscalar $\sigma$ and $\omega$ as well the isovector $\rho$ and $\delta$) and photon ($A$), the equation of motion for nucleons, i.e., the Dirac equation can be derived as,
\begin{equation}\label{equ:Dirac}
  [\bm{\alpha\cdot p}+\beta(M+\Sigma_S)+\Sigma_0]\psi=\varepsilon\psi,
\end{equation}
where $\Sigma_S$ and $\Sigma_0$ correspond to the scalar and vector potential, respectively, and $\varepsilon$ denotes the single-particle energy including the rest mass $M$. Here the calculations are restricted on the level of mean field approach with spherical symmetry, and the radial form of Eq. (\ref{equ:Dirac}) is then obtained as,
\begin{align}\label{equ:rDirac}
  \begin{pmatrix} M+\Sigma_+(r) & -\dfrac{d}{dr} + \dfrac{\kappa}{r}\\[0.5em] \dfrac{d}{dr} + \dfrac{\kappa}{r} & - M +\Sigma_-(r)  \end{pmatrix}\begin{pmatrix} g(r) \\[0.5em] f(r)  \end{pmatrix}= &\varepsilon\begin{pmatrix} g(r) \\[0.5em] f(r)  \end{pmatrix},
\end{align}
with $\Sigma_\pm(r)=\Sigma_0(r)\pm\Sigma_S(r)$ and $\kappa=(l-j)(2j+1)$. The radial wave functions $g(r)$ and $f(r)$ correspond to the upper and lower components of Dirac spinor, respectively.

Aiming at the full description on the half-lives of proton radioactivity within the CDF scheme, the key step is to determine the potential barrier compliant with the WKB approximation. However, the required potential barrier can not be extracted directly from Eq. (\ref{equ:rDirac}) due to the fact that the upper and lower components of the spinor are coupled. To solve this knotty problem, the SRG method is introduced to diagonalize the single-particle Hamiltonian \cite{Guo2012}, which leads to the decoupled form of Eq. (\ref{equ:rDirac}) as,
\begin{align}\label{equ:DH}
  \begin{pmatrix} H_1+M  & 0 \\[0.5em]  0 & H_2-M\end{pmatrix}\begin{pmatrix}  G(r) \\[0.5em] F(r)\end{pmatrix} = & \varepsilon \begin{pmatrix}  G(r) \\[0.5em] F(r)\end{pmatrix}.
\end{align}
$H_1$ and $G(r)$ stand for the single-particle Hamiltonian and radial wave function of nucleon in Fermi sea, respectively, which are exactly the required quantities in the following calculations, and $H_2$ and $F(r)$ are the ones for the antinucleon. Notice that the diagonalized Eq.(\ref{equ:DH}) consists of two independent Schr\"odinger-type equations for the Dirac particles and antiparticles respectively, since the upper wave function $G(r)$ and the lower one $F(r)$ have been decoupled. For the upper component of Eq. (\ref{equ:DH}) that describes the nucleons in Fermi sea, it can be expressed to the magnitude of $1/M^2$ as,
\begin{equation}\label{equ:rschr}
\begin{split}
  &\Big[-\frac{1}{2M}\frac{d^2}{dr^2}+\frac{l(l+1)}{2Mr^2}+\Sigma_+(r)-\frac{\kappa}{r}\frac{\Sigma_-^\prime}{4M^2}
      +\frac{\Sigma_+^{\prime\prime}}{8M^2}\\
      &-\Sigma_S\frac{l(l+1)}{2M^2r^2}+\frac{1}{2M^2}\Big(\Sigma_S\frac{d^2}{dr^2}+\Sigma_S^\prime\frac{d}{dr}\Big)\Big]G(r)
  =EG(r),
\end{split}
\end{equation}
where the single (double) prime denotes the first (second) derivative with respect to $r$ and single-particle energy $E=\varepsilon-M$ excluding the rest mass. The main parts of the potential in Eq.(\ref{equ:rschr}) are the $\Sigma_+(r)$ and centrifugal terms, i.e.,
\begin{equation}\label{equ:Vm}
  V_m(r) = \Sigma_+(r) + \frac{l(l+1)}{2Mr^2},
\end{equation}
and the fourth term in the square brackets corresponds to the spin-orbit coupling potential to the first order,
\begin{equation}\label{equ:Vso}
  V_{so}(r) = -\frac{\kappa}{r}\frac{\Sigma_-^\prime}{4M^2}.
\end{equation}
However, the potential in Eq. (\ref{equ:rschr}) contains the derivatives of the wave function, which induces non-locality and thus brings troubles in deducing the potential barrier. To overcome this difficulty, we replace the non-local term numerically by its equivalent local form, i.e.,  $\frac{1}{2M^2} \Big[\Sigma_SG''(r)+ \Sigma_S^\prime G'(r)\Big]/G(r)$. Thus the residual correction of the potential barrier, together with the rest terms, is labeled as,
\begin{align}\label{equ:Vrc}
  V_{rc}(r) = &\frac{\Sigma_+^{\prime\prime}}{8M^2}-\Sigma_S\frac{l(l+1)}{2M^2r^2} +\frac{\Sigma_SG''(r)+\Sigma_S^\prime G'(r)}{2M^2G(r)}.
\end{align}
Finally, the potential barrier we construct for the calculation of half-life of proton radioactivity can be written as,
\begin{equation}\label{equ:V}
  V(r)=V_m(r)+V_{so}(r)+V_{rc}(r),
\end{equation}
which is compliant with the WKB method.

\begin{figure}[htbp]
  \centering
  \includegraphics[width=0.47\textwidth]{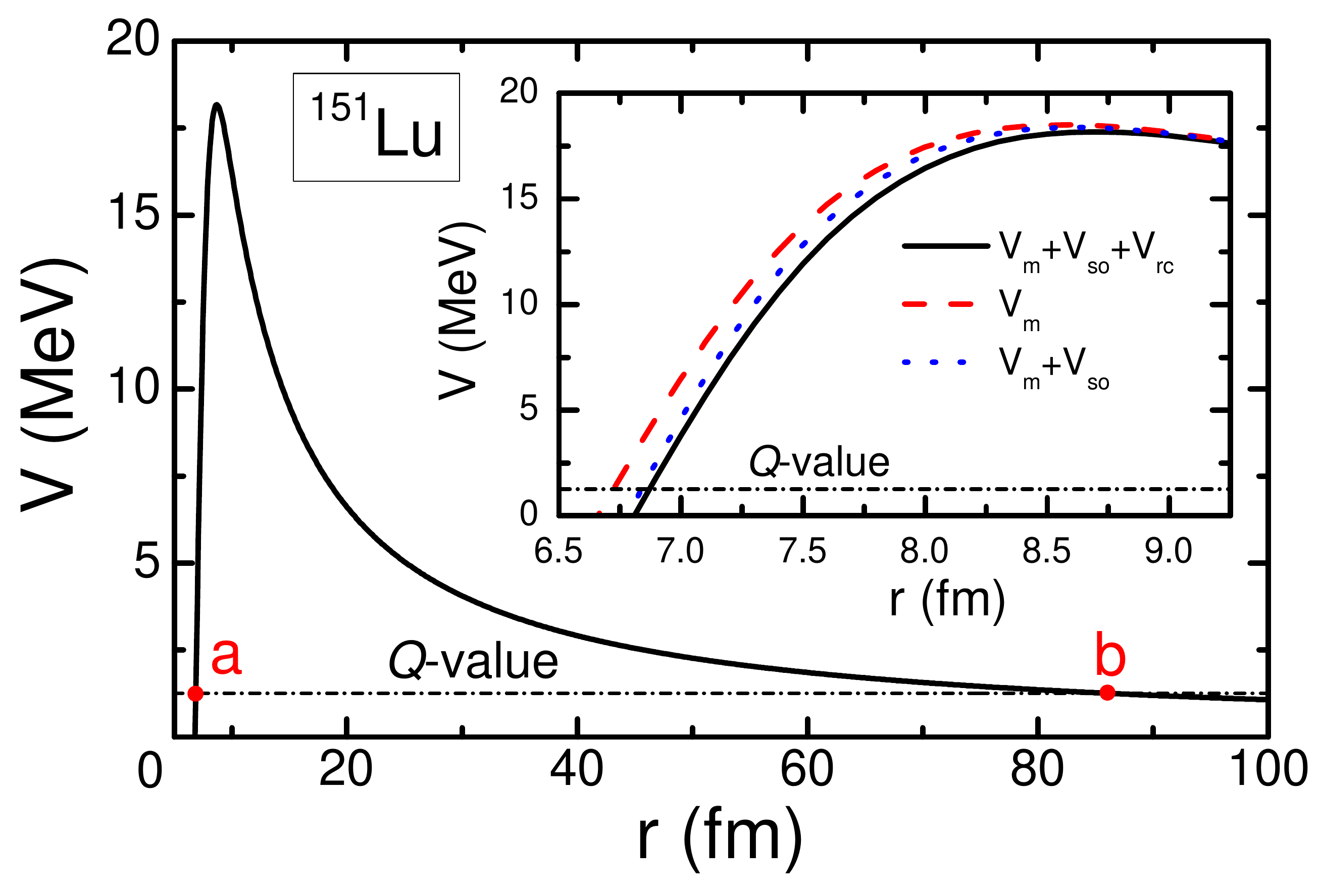}
  \caption{Potential barrier for the proton emission of $^{151}$Lu. The dash-dotted lines represent the $Q$-value, and $a$ and $b$ are the turning points. The inset shows the detailed contributions to the potential barrier defined in Eq. (\ref{equ:V}).}\label{fig:V}
\end{figure}

From Eq. (\ref{equ:V}), Fig. \ref{fig:V} illustrates the potential barrier $V(r)$ calculated with the CDF functional DD-ME$\delta$ \cite{Roca-Maza2011} by taking the proton emitter $^{151}$Lu as a candidate. It is shown that the top of the potential barrier is much higher than $Q$-value, which indicates that the WKB approximation remains valid here \cite{Hagino2004}. In determining the non-local contribution to residual correction term $V_{rc}$, the wave function $G(r)$ between two turning points $a$ and $b$ in Fig. \ref{fig:V} is approached with the WKB method as,
\begin{equation}\label{equ:G}
    G(r)=\frac{1}{\sqrt{v}}\exp\Big(-\Big|\int^b_aqdr\Big|\Big)\exp\Big(\Big|\int^r_bqdr\Big|\Big),
\end{equation}
where $q=\sqrt{2M[Q-V(r)]}$, $v=q/M$ and $Q$ is the decay energy. Considering the fact that the non-local contribution in $V(r)$ remains unknown here, the radial wave function $G(r)$ is calculated with the potential $V(r)$ that excludes the non-local term without loss of accuracy. Under the two-body scheme, the emitted proton moves through the potential barrier provided by the daughter nucleus and therefore the nucleon mass $M$ in Eqs. (\ref{equ:V}, \ref{equ:G}) should be replaced by the reduced one $\mu$. With the potential barrier calculated from Eq. (\ref{equ:V}), the half-life is then determined with following formula,
\begin{equation}\label{equ:T}
  T=\frac{\ln2}{\nu_0PS_p},
\end{equation}
where $\nu_0$, $S_p$ and $P$ denote the assault frequency, spectroscopic factor and barrier penetrability, respectively. As demonstrated in Ref. \cite{dong2011}, WKB approximation works well for proton radioactivity and its systematical deviations are compensated by the fitted assault frequency $\nu_0$. With the WKB approximation, the barrier penetrability $P$ is determined as,
\begin{equation}\label{penetration}
  P=\exp\Big\{{-2\int^b_a\sqrt{2\mu[V(r)-Q]}dr}\Big\},
\end{equation}
where the turning points $a$ and $b$ correspond with the radial positions of $V(r)=Q$ (see Fig. \ref{fig:V}).

As an important structural information of the single-particle levels around the Fermi surface, the spectroscopic factor $S_p$ is necessary to be introduced to improve the accuracy of the half-life calculations \cite{Dong2009}. In the case of proton radioactivity, it corresponds to the quantity $u^2_j$, i.e., the probability that the spherical orbit of the emitted proton is empty in the daughter nucleus \cite{Aberg1997,Delion2006}, which is extracted from the calculations of CDF + BCS. The pairing force is adopted as the density dependent zero-range force \cite{Bertsch1991}
\begin{equation}
  V_p(r_1,r_2)=V_0\delta(r_1-r_2)\Big[1-\frac{\rho_b(r)}{\rho_0}\Big],
\end{equation}
where $\rho_0$ is the saturation density and the pairing strength $V_0=-530$ MeV. Different from the neutron-rich side, the continuum effects can be taken into account reasonably by the BCS method with the pairing force above due to the existence of high barrier (see Fig. \ref{fig:V}) \cite{Lalazissis1998}. Thus in exploring the proton radioactivity it is appropriate to deal with the pairing correlations with the BCS method, which reduces the numerical task with equivalent accuracy \cite{Xiang2013}.

\begin{figure}[htbp]
  \centering
  \includegraphics[width=0.47\textwidth]{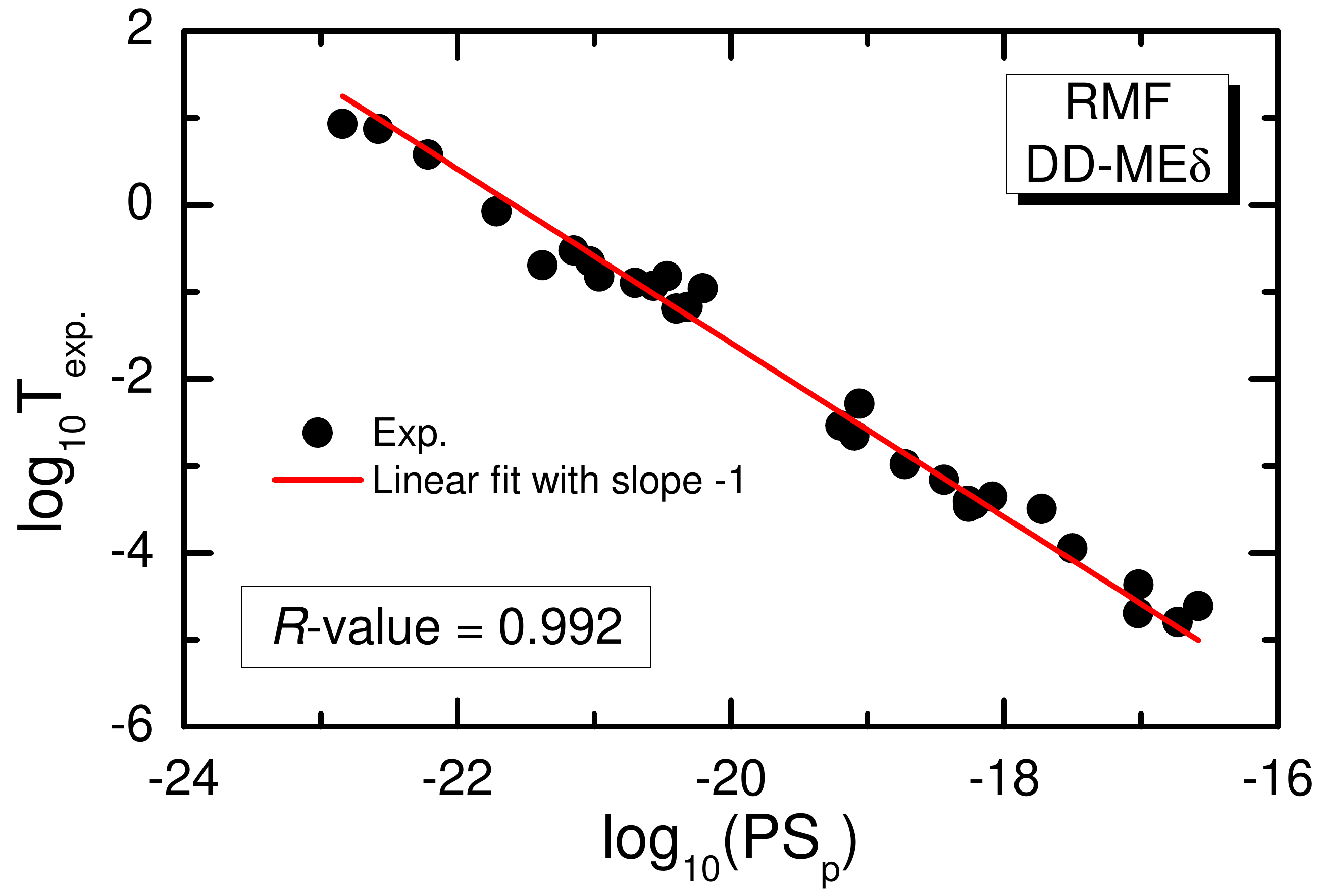}
  \caption{$\log_{10}T_{exp.}$ as a function of $\log_{10}(PS_p)$. All the 29 spherical proton emitters in Tab. \ref{tab:T} are used. The solid line is the fit line whose slope is limited to -1, and the fitted R-value is 0.992. }\label{fig:linear}
\end{figure}

Utilizing the CDF theory with DD-ME$\delta$ \cite{Roca-Maza2011}, the potential barrier is extracted with the SRG treatment. As the effective nuclear potential vanishes in a large distance, the radial truncation in CDF calculations is fixed to 30 fm. Table \ref{tab:T} lists the calculated spectroscopic factors and penetrability of proton radioactivity of 29 nearly spherical proton emitters,  in which the experimental $Q$-values and angular momentum transfer $l$ are also shown as the inputs to calculate the penetrability. The assault freuquency $\nu_0$ is assumed as a constant for all the proton emitters and determined by the linear relationship between the logarithms of $T$ and $PS_p$,
\begin{equation}\label{equ:logt}
  \log_{10}T=-\log_{10}(PS_p)+\log_{10}(\ln2)-\log_{10}\nu_0.
\end{equation}
With the experimental half-lives list in Tab. \ref{tab:T}, the assault frequency $\nu_0$ is fitted as $2.68 \times 10^{21}$s$^{-1}$. Figure \ref{fig:linear} displays the experimental $\log_{10}T_{exp.}$ of the selected 29 proton emitters with respect to the calculated $\log_{10}(PS_p)$, as well as the optimistic fitting. It is found that the linear relationship (\ref{equ:logt}) is fulfilled quite well, which is supported by the fact that the statistical correlation coefficient, namely, the $R$-value, is determined as large as 0.992. Thus the assumption of treating $\nu_0$ as a constant is confirmed as well, in coincidence with the estimations in Refs. \cite{dong0,dong1,Zhang2010}.

\renewcommand{\arraystretch}{1.1}
\begin{table*}[htbp]
  \centering
  \caption{Calculated and experimental half-lives of proton radioactivity for spherical proton emitters. Experiment $Q$ values, angular momentum transfer $l$ and half-lives are taken from Ref.\cite{Qi2012}. The calculated half-lives of proton radioactivity $T_{NL}^{\rm BCS}$ and $T_{\rm R3Y}$ in Ref. \cite{Ferreira2011,Sahu2011} are presented. The calculated errors of $T_{cal.}$ are induced by the experiment error on the $Q$ value. The last column is the relative deviations (RD) of $|T_{cal.}-T_{exp.}|/T_{exp.}\times100\%$. An asterisk ($*$) denotes the isomeric state.}\label{tab:T}
  \begin{ruledtabular}
    \begin{tabular}{cccccrrcccrr}
    Emitter        & $l$ & $Q$(keV) &          $P$          & $S_p$ & $T_{exp.}$                        &  $T_{cal.}$                                   & $T_{NL}^{\rm BCS}$ \cite{Ferreira2011}& $T_{\rm R3Y}$ \cite{Sahu2011}& $C_{so}$  & $C_{rc}$  &  $RD     $ \\ \hline
    $^{146}$Tm~    &  5  & 1210(4)  & $3.732\times10^{-21}$ & 0.728 & 117.6$(^{+6.4  }_{-6.4  })    $ms & {\bf 95.4 $\bm{(^{+10.2 }_{-9.2  })    }$ ms} &                                       & 88.1 s                          & 20.8\%    & 12.9\%    & ~18.9\%      \\ 
    $^{146}$Tm$^*$ &  5  & 1140(4)  & $5.746\times10^{-22}$ & 0.728 & 203  $(^{+6    }_{-6    })    $ms &      619  $    (^{+73   }_{-65   })     $ ms  &                                       & 328 ms                          & 21.0\%    & 13.0\%    & 205.0\%      \\ 
    $^{147}$Tm~    &  5  & 1073(5)  & $8.422\times10^{-23}$ & 0.725 & 3.78 $(^{+1.27 }_{-1.27 })    $ s & {\bf 4.24 $\bm{(^{+0.70 }_{-0.60 })    }$  s} & 2.5 s                                 & 595.7 s                         & 19.6\%    & 12.5\%    & ~12.3\%      \\ 
    $^{147}$Tm$^*$ &  2  & 1133(3)  & $9.920\times10^{-19}$ & 0.607 & 0.360$(^{+0.036}_{-0.036})    $ms & {\bf 0.430$\bm{(^{+0.038}_{-0.034})    }$ ms} & 0.20 ms                               & 0.28 ms                         & -3.2\%    & ~8.8\%    & ~19.4\%      \\ 
    $^{150}$Lu~    &  5  & 1283(3)  & $6.520\times10^{-21}$ & 0.611 & 64.0 $(^{+5.6  }_{-5.6  })    $ms & {\bf 65.1 $\bm{(^{+4.9  }_{-4.5  })    }$ ms} &                                       & 2.29 s                          & 20.0\%    & 11.4\%    & ~~1.6\%      \\ 
    $^{150}$Lu$^*$ &  2  & 1306(5)  & $1.936\times10^{-17}$ & 0.495 & 43   $(^{+7    }_{-5    })~\mu$s  &      27   $    (^{+3    }_{-3    })~\mu $s    &                                       & 9.16 $\mu$s                     & -3.3\%    & ~8.6\%    & ~37.1\%      \\ 
    $^{151}$Lu~    &  5  & 1253(3)  & $3.252\times10^{-21}$ & 0.614 & 127.1$(^{+1.8  }_{-1.8  })    $ms & {\bf 129.8$\bm{(^{+10.0 }_{-9.4  })    }$ ms} & 70 ms                                 & 4.94 s                          & 19.9\%    & 11.0\%    & ~~2.1\%      \\ 
    $^{151}$Lu$^*$ &  2  & 1332(10) & $3.612\times10^{-17}$ & 0.514 & 16   $(^{+1    }_{-1    })~\mu$s  & {\bf 14   $\bm{(^{+4    }_{-3    })~\mu}$s  } & 7.2 $\mu$s                            & 5.93 $\mu$s                     & -3.8\%    & ~8.4\%    & ~12.8\%      \\ 
    $^{155}$Ta~    &  5  & 1468(15) & $1.266\times10^{-19}$ & 0.499 & 2.9  $(^{+1.5  }_{-1.1  })    $ms & {\bf 4.1  $\bm{(^{+1.5  }_{-1.1  })    }$ ms} & 2.5 ms                                & 57.8 $\mu$s                     & 19.3\%    & 12.8\%    & ~41.4\%      \\ 
    $^{156}$Ta~    &  2  & 1032(5)  & $2.418\times10^{-21}$ & 0.452 & 149  $(^{+8    }_{-8    })    $ms &      237  $    (^{+44   }_{-37   })     $ ms  &                                       & 158 ms                          & -3.2\%    & 10.0\%    & ~59.1\%      \\ 
    $^{156}$Ta$^*$ &  5  & 1127(7)  & $2.983\times10^{-23}$ & 0.487 & 8.52 $(^{+2.12 }_{-2.12 })    $ s &      17.85$    (^{+4.24 }_{-3.36 })     $  s  &                                       & 1084 s                          & 19.3\%    & 12.6\%    & 109.5\%      \\ 
    $^{157}$Ta~    &  0  & ~~947(7) & $9.037\times10^{-22}$ & 0.785 & 0.300$(^{+0.105}_{-0.105})    $ s & {\bf 0.365$\bm{(^{+0.113}_{-0.086})    }$  s} & 0.23 s                                & 0.104 s                         & ~1.3\%    & ~7.8\%    & ~21.7\%      \\ 
    $^{159}$Re$^*$ &  5  & 1831(20) & $2.544\times10^{-17}$ & 0.374 & 20.2 $(^{+3.7  }_{-3.7  })~\mu$s  & {\bf 27.2 $\bm{(^{+9.6  }_{-7.0  })~\mu}$s  } & 18 $\mu$s                             &                                 & 19.5\%    & 11.0\%    & ~34.9\%      \\ 
    $^{160}$Re~    &  2  & 1287(6)  & $9.615\times10^{-19}$ & 0.377 & 0.687$(^{+0.011}_{-0.011})    $ms & {\bf 0.715$\bm{(^{+0.116}_{-0.099})    }$ ms} &                                       & 0.250 ms                        & -3.8\%    & ~8.9\%    & ~~4.1\%      \\ 
    $^{161}$Re~    &  0  & 1214(6)  & $1.100\times10^{-18}$ & 0.743 & 0.440$(^{+0.002}_{-0.002})    $ms &      0.317$    (^{+0.056}_{-0.047})     $ ms  & 0.19 ms                               & 0.082 ms                        & ~1.3\%    & ~7.3\%    & ~28.0\%      \\ 
    $^{161}$Re$^*$ &  5  & 1338(6)  & $2.585\times10^{-21}$ & 0.361 & 224  $(^{+31   }_{-31   })    $ms & {\bf 278  $\bm{(^{+43   }_{-37   })    }$ ms} & 0.20 s                                & 5.13 s                          & 18.6\%    & 12.8\%    & ~24.0\%      \\ 
    $^{164}$Ir~    &  5  & 1844(9)  & $1.271\times10^{-17}$ & 0.248 & 0.113$(^{+0.062}_{-0.030})    $ms & {\bf 0.082$\bm{(^{+0.012}_{-0.011})    }$ ms} &                                       & 0.166 ms                        & 18.6\%    & 11.3\%    & ~27.1\%      \\ 
    $^{165}$Ir$^*$ &  5  & 1733(7)  & $2.265\times10^{-18}$ & 0.241 & 0.34 $(^{+0.07 }_{-0.07 })    $ms &      0.48 $    (^{+0.06 }_{-0.05 })     $ ms  & 0.41 ms                               & 1.22 ms                         & 18.6\%    & 11.2\%    & ~39.9\%      \\ 
    $^{166}$Ir~    &  2  & 1168(7)  & $1.042\times10^{-20}$ & 0.329 & 0.152$(^{+0.071}_{-0.071})    $ s & {\bf 0.076$\bm{(^{+0.017}_{-0.014})    }$  s} &                                       & 0.029 s                         & -3.2\%    & ~9.1\%    & ~50.2\%      \\ 
    $^{166}$Ir$^*$ &  5  & 1340(8)  & $8.499\times10^{-22}$ & 0.228 & 0.84 $(^{+0.28 }_{-0.28 })    $ s & {\bf 1.34 $\bm{(^{+0.29 }_{-0.24 })    }$  s} &                                       & 15.2 s                          & 18.7\%    & 11.6\%    & ~59.0\%      \\ 
    $^{167}$Ir~    &  0  & 1096(6)  & $8.245\times10^{-21}$ & 0.758 & 110  $(^{+15   }_{-15   })    $ms &      42   $    (^{+9    }_{-7    })     $ ms  & 41 ms                                 & 15.9 ms                         & ~1.3\%    & ~8.0\%    & ~62.3\%      \\ 
    $^{167}$Ir$^*$ &  5  & 1261(7)  & $1.194\times10^{-22}$ & 0.222 & 7.5  $(^{+2.4  }_{-2.4  })    $ s & {\bf 9.8  $\bm{(^{+2.0  }_{-1.7  })    }$  s} & 7.5 s                                 & 150 s                           & 17.8\%    & 13.1\%    & ~30.4\%      \\ 
    $^{170}$Au~    &  2  & 1488(12) & $8.383\times10^{-18}$ & 0.224 & 321  $(^{+67   }_{-58   })~\mu$s  &      138  $    (^{+40   }_{-31   })~\mu $s    &                                       &                                 & -3.1\%    & 10.2\%    & ~57.0\%      \\ 
    $^{170}$Au$^*$ &  5  & 1770(6)  & $1.623\times10^{-18}$ & 0.115 & 1.046$(^{+0.136}_{-0.126})    $ms &      1.384$    (^{+0.144}_{-0.131})     $ ms  &                                       &                                 & 17.9\%    & 11.6\%    & ~32.3\%      \\ 
    $^{171}$Au~    &  0  & 1464(10) & $3.494\times10^{-17}$ & 0.747 & 24.5 $(^{+4.7  }_{-3.1  })~\mu$s  &      9.9  $    (^{+2.4  }_{-1.9  })~\mu $s    & 6.8 $\mu$s                            & 2.29 $\mu$s                     & ~1.4\%    & ~7.6\%    & ~59.5\%      \\ 
    $^{171}$Au$^*$ &  5  & 1719(4)  & $7.130\times10^{-19}$ & 0.112 & 2.22 $(^{+0.19 }_{-0.19 })    $ms &      3.24 $    (^{+0.23 }_{-0.22 })     $ ms  & 3.1 ms                                & 4.53 ms                         & 17.7\%    & 11.4\%    & ~45.7\%      \\ 
    $^{176}$Tl~    &  0  & 1282(18) & $1.238\times10^{-19}$ & 0.702 & 5.2  $(^{+3.0  }_{-1.4  })    $ms & {\bf 3.0  $\bm{(^{+1.9  }_{-1.1  })    }$ ms} &                                       &                                 & ~1.3\%    & ~8.2\%    & ~42.7\%      \\ 
    $^{177}$Tl~    &  0  & 1180(20) & $6.789\times10^{-21}$ & 0.712 & 67   $(^{+37   }_{-37   })    $ms & {\bf 54   $\bm{(^{+46   }_{-24   })    }$ ms} & 48 ms                                 & 11.9 ms                         & ~1.3\%    & ~7.7\%    & ~20.0\%      \\ 
    $^{177}$Tl$^*$ &  5  & 1984(8)  & $1.965\times10^{-17}$ & 0.028 & 396  $(^{+87   }_{-77   })~\mu$s  & {\bf 478  $\bm{(^{+59   }_{-51   })~\mu}$s  } & 234 $\mu$s                            & 66.4 $\mu$s                     & 16.9\%    & 11.0\%    & ~20.6\%      \\
  \end{tabular}
  \end{ruledtabular}
\end{table*}

With the fitted assault frequency $\nu_0$, the half-lives of proton radioactivity $T_{cal.}$ are then determined by Eq. (\ref{equ:T}) with the potential barrier (\ref{equ:V}) and the results are listed in Tab. \ref{tab:T}. For comparison, another two theoretical calculations are also presented, namely, $T_{NL}^{\rm BCS}$ \cite{Ferreira2011} and $T_{\rm R3Y}$ \cite{Sahu2011}. In Ref. \cite{Sahu2011}, the potential barrier was determined by the single folding model with a microscopic interaction R3Y that was derived from the linear CDF theory, and the half-lives are calculated by the WKB approximation but without including the spectroscopic factor. For most of the selected emitters, it can reproduce the experimental half-lives properly. However, for some emitters, the deviations from the data are on several orders of magnitude, such as, $^{146}$Tm, $^{147}$Tm, $^{155}$Ta and $^{156}$Ta$^*$. It is mentioned in Ref. \cite{Sahu2011} that further correction for the R3Y effective interaction is essential. Base on the scattering theory which starts from a Schr\"odinger equation, Ferreira {\it et al.} calculated the half-lives within the non-linear CDF scheme \cite{Ferreira2011}, and the results ($T_{NL}^{BCS}$) agree with the data within one order of magnitude for their selected odd emitters.

Compared to $T_{NL}^{BCS}$ and $T_{R3Y}$, our calculations $T_{cal.}$ show much better agreement with the data than $T_{R3Y}$ and similar quantitative accuracy as $T_{NL}^{BCS}$ for most of the emitters. As shown in Tab. \ref{tab:T}, the ratios of the calculated half-life $T_{cal.}$ over the experimental one $T_{exp.}$ are found within the range from $1/3$ to $3$. Specifically for most of the emitters the relative deviations (RD) of $T_{cal.}$ from the data are less than $50\%$ as seen from the last column of Tab. \ref{tab:T}, correspondingly the ratio $T_{cal.}/T_{exp.}$ lying within the range from $0.5$ to $1.5$. In addition, for 18 of 29 selected emitters, the calculated half-lives can reproduce within the range of the experimental and theoretical error bars (denoted in bold type) and the later originate from the uncertainties in the $Q$-values. On the one hand, it suggests that the present approach can be used not only to estimate the half-lives of proton radioactivity, but also in turn to extract the structural information of emitters combined with experimental measurements. For example, with the decay energy $Q=1.468$ MeV and $T=2.9^{+1.5}_{-1.1}$ ms for the new proton emitter $^{155}$Ta, the angular momentum transfer $l$ can be determined theoretically as $l=4.9$ and therefore the proton is emitted from $\pi h_{11/2}$ orbit which agrees with the conclusion in Ref. \cite{Joss2006}. On the other hand, the agreement between the theoretical calculations and experimental data indicates the potential barriers as well as spectroscopic factor $S_p$ from the CDF calculations are reasonable, which suggests the CDF models can be applied to describe the proton-rich nuclei that far away from the $\beta$-stable line to a large extent. Large discrepancies are found between theoretical and experimental results for $^{146}$Tm$^*$ and $^{156}$Ta$^*$. For the former, if the experimental $Q$-value is adopted as 1.199 MeV from AME2012 \cite{Audi2012}, the half-life will be 127 ms which is much closer to the experiment value 203 ms than the one with the $Q$-value 1.140 MeV in Tab. \ref{tab:T}. It seems that 1.199 MeV is more reliable as the Q-value for the emitter $^{146}$Tm$^*$. While for $^{156}$Ta$^*$, if we select the $Q$-value as 1.114 MeV from AME2012 that is smaller than the latest data 1.127 MeV \cite{Darby2011}, a longer half-life than 17.85 s is obtained, deviating further away from the measured value 8.52 s.

The potential barrier extracted from the radial Dirac equation (\ref{equ:rDirac}) with the SRG method consists of three parts, namely the main part $V_m$, spin-orbit term $V_{so}$ and residual correction $V_{rc}$ [see Eqs. (\ref{equ:Vm}-\ref{equ:Vrc})]. Taking the emitter $^{151}$Lu as an example, the potentials $V_{m}$, $V_m+V_{so}$ and $V_m+V_{so}+V_{rc}$ are plotted in the inset of Fig. \ref{fig:V}. It is obvious that the potential barrier is mainly contributed by $V_m$, the mean potential $\Sigma_+$ plus centrifugal barrier. Even though, the relativistic corrections to the barrier, i.e., the spin-orbit term $V_{so}$ and residual correction $V_{rc}$, are still essential due to the fact that the penetrability exponentially depends on the barrier [see Eq. (\ref{penetration})]. To quantify the effects of the spin-orbit potential and residual correction term, we evaluate the half-lives respectively with potentials $V_m+V_{so}$ and $V_m+V_{rc}$ , namely $T_{cal.}^{V_m+V_{so}}$ and $T_{cal.}^{V_m+V_{rc}}$, without changing the assault frequency $\nu_0$. Their relative contributions in the half-lives listed in Tab. \ref{tab:T} are defined as
\begin{align}
  C_{so} &= (T_{cal.}^{V_m+V_{rc}}-T_{cal.})/T_{cal.}\times100\%,\\
  C_{rc} &= (T_{cal.}^{V_m+V_{so}}-T_{cal.})/T_{cal.}\times100\%.
\end{align}
The contribution of $V_{so}$ can be as large as 21.0\% while for the lower angular momentum transfer, e.g., $l= 0$ and 2, the the effect becomes weaker, with $C_{so}$ being less than 4\%. The contribution of $V_{rc}$ that can be as large as 13\% is also found to be related to the angular momentum transfer while not so distinct as $V_{so}$.

To further confirm the effects of $V_{so}$ and $V_{rc}$, we introduce the root mean square deviation (RMSD) of the logarithm of the theoretical half-lives from the experimental one
\begin{equation}\label{equ:rmsd}
  \text{RMSD}=\sqrt{\sum^{29}_{i=1}\frac{(\log_{10}T_{cal.,\ i}-\log_{10}T_{exp.,\ i})^2}{29}}.
\end{equation}
Here the assault frequency is optimized for each following choice of the barrier. When $V_{so}$ and $V_{rc}$ are excluded from the potential barrier $V(r)$, i.e., only with potential $V_m$, the RMSD value is found to be 0.242. Yet, when the potential $V_m$ is implemented with $V_{so}$ ($V_{rc}$), the RMSD value can be reduced to 0.215 (0.236), which suggests that the role of $V_{so}$ ($V_{rc}$) is non-negligible for the accurate description of the half-lives. With both terms ($V_{so}$ and $V_{rc}$) included in the barrier $V(r)$, the RMSD value can be further reduced to 0.211. In addition, we also find that the higher order terms (than $1/M^2$) in the barrier can be completely neglected in calculating the half-lives.

On the other hand, RMSD value can also be used to quantify the significance of the spectroscopic factor in determining the half-lives. Without considering the spectroscopic factor, the RMSD value is 0.349. However, when it is included, the RMSD value is distinctly reduced as 0.211. As pointed out in Ref. \cite{Dong2009}, such improvement is due to the fact that the spectroscopic factor contains the shell effect and other structural information that makes the description of proton radioactivity more accurate and reliable.

In summary, the proton radioactivity of spherical proton emitters has been studied under the framework of covariant density functional (CDF) theory as combined with the WKB approximation, and for the first time the potential barrier that prevents the emitted proton is extracted from the radial Dirac equation with the similarity renormalization group (SRG) method. With SRG treatment, the relativistic corrections in potential barrier, namely the spin-orbit potential and residual correction term can be deduced naturally from the non-relativistic reduction of Dirac equation, which present distinct effects in determining the half-lives of proton radioactivity. The spectroscopic factor determined by the self-consistent calculation of CDF + BCS model is also taken into account, and its significance in describing the half-lives of proton radioactivity is manifested once again in terms of the RMSD value. As an extended application of the SRG method within the CDF scheme, the current approach well reproduces the experimental data of the half-life, which may indicate the reliability of the CDF theory in describing the proton-rich nuclei.

This work is partly supported by the National Natural Science Foundation of China under Grant No. 11375076, the Specialized Research Fund for the Doctoral Program of Higher Education under Grant No. 20130211110005, and the Youth Innovation Promotion Association of Chinese Academy of Sciences.

\end{document}